\documentclass[prl,aps,floats,twocolumn]{revtex4}

\usepackage[dvips]{graphicx}
\usepackage{amssymb}
\begin{document}

\title{The Cosmic 
Gravitational-Wave Background in a Cyclic Universe}
\author{Latham A. Boyle$^1$, Paul J. Steinhardt$^1$, and Neil Turok$^2$}
\address{$^1$Department of Physics, Princeton University, Princeton,
New Jersey 08544 \\
$^2$Department of Applied Mathematics and Theoretical Physics,
Centre for Mathematical Sciences, University of Cambridge, Wilberforce
Road, Cambridge CB3 OWA, United Kingdom}
\date{July 2003}

\begin{abstract}
Inflation predicts a primordial gravitational wave spectrum that is
slightly ``red,'' {\it i.e.} nearly scale-invariant with slowly
increasing power at longer wavelengths.  In this paper, we compute
both the amplitude and spectral form of the primordial tensor spectrum
predicted by cyclic/ekpyrotic models.  The spectrum is exponentially
suppressed compared to inflation on long wavelengths, and the
strongest constraint emerges from the requirement that the energy
density in gravitational waves should not exceed around $10$ per cent
of the energy density at the time of nucleosynthesis.
\end{abstract}
\maketitle

The recently-proposed cyclic model\cite{cyclic_intro,cyclic_evolution}
differs radically from standard inflationary cosmology
\cite{old_inflation,new_inflation}, while retaining the inflationary
predictions of homogeneity, flatness, and nearly scale-invariant
density perturbations.  It has been suggested that the cosmic
gravitational wave background provides the best experimental means for
distinguishing the two models.  Inflation predicts a nearly
scale-invariant (slightly red) spectrum of primordial tensor
perturbations, whereas the cyclic model predicts a blue
spectrum.\cite{cyclic_intro} The difference arises because inflation
involves an early phase of hyper-rapid cosmic acceleration, whereas
the cyclic model does not.

In this paper, we compute the gravitational wave spectrum for cyclic
models to obtain both the normalization and spectral shape as a
function of model parameters, improving upon earlier heuristic
estimates.  We find that the spectrum is strongly blue. The amplitude
is too small to be observed by currently proposed detectors on all
scales. Hence, the discovery of a stochastic background of
gravitational waves would be evidence in favor of inflation, and
would rule out the cyclic model.

\begin{figure}
\begin{center}
\includegraphics[width=3.0in]{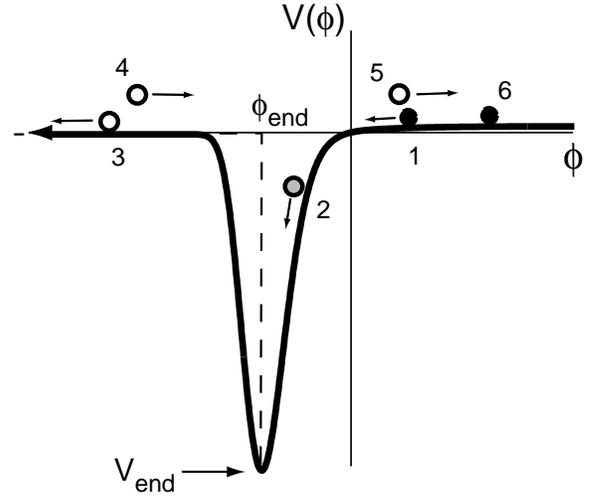}
\end{center}
\caption{Schematic of cyclic potential with numbers representing the
stages described in the text.  To the left of $\phi_{end}$, where the
scalar kinetic energy dominates, we approximate $V$ with a Heaviside
function, jumping to zero as shown by the dashed line.  }
\label{pot}
\end{figure}

Readers unfamiliar with the cyclic model may consult
\cite{cyclic_informal} for an informal tour, and
\cite{cyclic_constraints} for a recent analysis of phenomenological
constraints.  Cyclic cosmology draws strongly on earlier ideas
associated with the ``ekpyrotic universe''
scenario. \cite{ekpyrosis,crunch_bang,ek_density_pert} Briefly, the
scenario can be described in terms of the periodic collision of
orbifold planes moving in an extra spatial dimension, or,
equivalently, in terms of a four-dimensional theory with an evolving
(modulus) field $\phi$ rolling back and forth in an effective
potential $V(\phi)$.  The field theory description is the long
wavelength approximation to the brane picture in which the potential
represents the interbrane interaction and the modulus field determines
the distance between branes.  For the purposes of this paper, the
field theoretic description is more useful.

The potential (Fig.~1) is small and positive for large $\phi$, falling
steeply negative at intermediate $\phi$, and increasing again for
negative $\phi$. Each cycle consists of the following stages: (1)
$\phi$ large and decreasing: the universe expands at an accelerated
rate as $V(\phi)>0$ acts as dark energy; (2) $\phi$ intermediate and
decreasing: the universe is dominated by a combination of scalar
kinetic and potential energy, leading to slow contraction and to the
generation of fluctuations; (3) $\phi$ negative and decreasing
(beginning at conformal time $\tau_{end}<0$): the generation of
fluctuations ends, $\phi$ rolls past $\phi_{end}$ and, in the
four-dimensional description, the universe contracts rapidly,
dominated by scalar field kinetic energy, to the bounce ($\tau=0$) at
which matter and radiation are generated; (4) $\phi$ increasing from
minus infinity: the universe remains dominated by scalar field kinetic
energy, which decreases rapidly compared to the radiation energy; (5)
$\phi$ large and increasing (beginning at $\tau_r>0$): the scalar
field kinetic energy red-shifts to a negligible value and the universe
begins the radiation dominated expanding phase; (6) $\phi$ large and
nearly stationary: the universe undergoes the transitions to matter
and dark energy domination, and the cycle begins anew.

We model the scalar field potential as:
\begin{equation}
\label{sudden_potential}
V(\phi) = V_{0}(1-e^{-c\phi/M_{pl}})\Theta(\phi-\phi_{end})
\end{equation}
where $M_{pl}$ is the reduced Planck mass and $\Theta(\phi)$ the
Heaviside step function. A potential of this form, with an
exponentially steep form, is required by the cyclic model in order to
produce an acceptable spectrum of cosmological perturbations.
\cite{cyclic_evolution,cyclic_constraints} Choosing $c=10$ for
example results in a scalar spectral index $n_{s}=.96$ which is
compatible with current constraints.  The Heaviside function marks the
end of the steeply decreasing part of the potential; for $\phi<
\phi_{end}$ the potential is small and the universe is dominated by
scalar field kinetic energy.

Our calculation begins in the ``ekpyrotic phase,'' stage (2), with the
Einstein-frame scale factor contracting:
\begin{equation}
\label{a(tau)_ek}
a(\tau) = a_{end}
  \left( \frac{\tau - \tau_{ek}}{\tau_{end} - \tau_{ek}} \right)^{\alpha},
  \qquad \tau<\tau_{end}\;,
\end{equation}
where $\alpha\equiv 2/(c^{2}-2)\ll 1$ and
$\tau_{ek}\equiv(1-2\alpha)\tau_{end}$, being the conformal time the
potential would have diverged to minus infinity had the exponential
form continued.  At $\tau=\tau_{end}$, the ekpyrotic phase ends and
the ``contracting kinetic phase,'' stage (3), begins:
\begin{equation}
\label{a(tau)_contracting_kinetic}
 a(\tau)=\left(\frac{-\;\tau}{(1+\chi)\tau_{r}}\right)^{1/2},
  \qquad \tau_{end}<\tau<0\;.
\end{equation}
At $\tau=0$, the universe bounces and the ``expanding kinetic phase,''
stage (4), begins:
\begin{equation}
\label{a(tau)_expanding_kinetic}
 a(\tau) = \left(\frac{\tau}{\tau_{r}}\right)^{1/2},
  \qquad 0<\tau<\tau_{r}\;.
\end{equation}
Radiation is produced at the bounce, but is less than the scalar
kinetic energy until, at $\tau=\tau_{r}$, the expanding kinetic phase
ends, and standard radiation-dominated, matter-dominated, and
dark-energy-dominated epochs ensue.  The transition times, $\tau_{r}$
and $\tau_{end}$, are given by
\begin{equation}
\label{tau_r}
\tau_{r} = (\sqrt{2}H_{r})^{-1}, \qquad
\tau_{end} = -\;\tau_{r}/\Gamma,
\end{equation}
and
\begin{equation}
\label{def_Gamma}
\Gamma\equiv\left|\frac{\tau_{r}}{\tau_{end}}\right|
=\left[\frac{1}{1+\chi}\left(\frac{2\alpha}{1-2\alpha}\right)\left(
\frac{V_{end}}{H_{r}^{2}M_{pl}^{2}}\right)\right]^{1/3},
\end{equation}
where $H_{r}\equiv H(\tau_{r})$ is the Hubble constant at $\tau_{r}$,
$V_{end}= -V(\phi_{end})$ is the depth of the potential at its
minimum, and $\chi\ll 1$ is a small positive constant that measures
the amount of radiation created at the bounce.  Note that $a(\tau)$
and $a'(\tau)$ are both continuous at the transition time
$\tau=\tau_{end}$, and we have chosen to normalize $a(\tau)$ to unity
at the start of radiation domination ($a(\tau_{r})=1$).

\section{The primordial spectrum, $\Delta h(k,\tau_{r})$}

A quasi-stationary, isotropic, stochastic background of gravitational
waves is characterized by the quantity $\Delta h(k,\tau)$, the rms
dimensionless strain per unit logarithmic wavenumber at time $\tau$
(i.e. the $\delta L/L$ that would be measured by a detector with
sensitivity band centered on mode $k$ and bandwidth $\Delta k=
k$). Accounting for both polarizations, it is given by $\Delta
h(k,\tau)=k^{3/2}|h_{k}(\tau)|/\pi$, where the Fourier amplitude
$h_{k}(\tau)$ satisfies
\begin{equation}
\label{h_EOM}
h_{k}''+2\frac{a'}{a}h_{k}'+k^{2}h_{k}=0
\end{equation}
and the boundary condition that the solution approaches the Minkowski
vacuum at short distances
\begin{equation}
\label{h_past}
h_{k}(\tau)\rightarrow\frac{e^{-ik\tau}}{a(\tau)M_{pl}\sqrt{2 k}}\;
\quad\textrm{as}\quad\tau\rightarrow-\infty.
\end{equation}

To solve equation (\ref{h_EOM}), it is useful to define $f_{k}(\tau)
\equiv a(\tau)\,h_{k}(\tau)$ and rewrite (\ref{h_EOM}) as
\begin{equation}
\label{f_EOM}
(f_{k})'' + (k^{2} - \frac{a''}{a})f_{k} = 0\;.
\end{equation}
During the ekpyrotic phase, $a(\tau)$ is given by (\ref{a(tau)_ek}),
and the general solution of (\ref{f_EOM}) is
\begin{equation}
\label{exact_ek_soln}
f_{k}(\tau)=\sqrt{y}\left(A_{1}(k)H_{n}^{(1)}(y)+A_{2}(k)H_{n}^{(2)}(y)\right),
\end{equation}
where $A_{1,2}(k)$ are arbitrary constants, $n\equiv{1\over
2}-\alpha$, $y\equiv -k(\tau-\tau_{ek})$, and $H_{n}^{(1,2)}$ are the
Hankel functions.  The boundary condition (\ref{h_past}) implies
\begin{equation}
\label{A_k}
A_{1}(k)=\frac{1}{2}\sqrt{\frac{\pi}{k}}, \qquad A_{2}(k)=0,
\end{equation}
where we have dropped a physically irrelevant phase.  In the
contracting kinetic phase, stage (4), $a(\tau)$ is given by
(\ref{a(tau)_contracting_kinetic}), and the general solution of
(\ref{f_EOM}) is
\begin{equation}
\label{exact_contracting_kinetic_soln}
\!f_{k}(\tau)\!=\!\sqrt{-k\tau}\!\left(\!B_{1}(k)H_{0}^{(1)}\!(-k\tau)
\!+\!B_{2}(k)H_{0}^{(2)}\!(-k\tau)\!\right)
\end{equation}
where $B_{1,2}(k)$ are arbitrary constants.  Then, continuity of
$h_{k}$ and $h_{k}'$ at $\tau=\tau_{end}$ implies
\begin{eqnarray}
\label{B_k}
B_{1,2}(k) & = & \mp\frac{i\pi}{4}\sqrt{\frac{\pi\alpha}{2k}}x_{e}
  \left[H_{1}^{(2,1)}(x_{e})H_{n}^{(1)}(2\alpha x_{e})+
  \right. \nonumber \\
 & & \left. +H_{0}^{(2,1)}(x_{e})H_{n-1}^{(1)}(2\alpha x_{e})\right]
\end{eqnarray}
where $x_{e}\equiv k|\tau_{end}|$.  Finally, in the expanding kinetic
phase, $a(\tau)$ is given by (\ref{a(tau)_expanding_kinetic}), and the
general solution of (\ref{f_EOM}) is
\begin{equation}
\label{exact_expanding_kinetic_soln}
f_{k}(\tau)=\sqrt{k\tau}\left(C_{1}(k)H_{0}^{(1)}(k\tau)
+C_{2}(k)H_{0}^{(2)}(k\tau)\right).
\end{equation}
To fix $C_{1,2}(k)$, we need to match the solution across $\tau=0$.
At the level of quantum field theory in curved space-time, the choice
is essentially unique\cite{Tolley_Turok}, and amounts to analytically
continuing the positive (negative) frequency part of
$h_{k}\!\equiv\!f_{k}/a$ around the origin in the lower (upper) half
of the complex $\tau$-plane, so $H_{0}^{(1,2)}(-k\tau)\!\rightarrow\!
-H_{0}^{(2,1)}(k\tau)$.  This yields
\begin{equation}
\label{C_k}
C_{1,2}(k)=-\sqrt{1+\chi}\;B_{2,1}(k)\;.
\end{equation}
The pre-factor arises because $a(\tau)$ differs by a factor of $\sqrt{1
+\chi}$ between the kinetic contracting and expanding phases; see
Eqs.~(\ref{a(tau)_contracting_kinetic}) and
(\ref{a(tau)_expanding_kinetic}).  Combining our results, we arrive at
the ``primordial'' dimensionless strain spectrum at the beginning of
the radiation dominated epoch:
\begin{equation}
\label{primordial_strain}
\begin{array}{ll}
\Delta h(k,\tau_{r})= & \;(k^{2}/\pi M_{pl})
\;\sqrt{2(1+\chi\,)\tau_{r}}\nonumber \\
& \left|B_{2}(k)H_{0}^{(1)}(x_{r})+ B_{1}(k)H_{0}^{(2)}(x_{r})\right|
\end{array}
\end{equation}
where $x_{r}\equiv k\tau_{r}$ and $k< k_{end}$.
For $k>k_{end}$, the spectrum is cut off because these modes
are not amplified and, instead, Eq.~(\ref{primordial_strain})
converges to the result for a static Minkowski background.
Recall, in the string theory context, we are
describing the collision of two orbifold planes and in the vicinity of
the collision the background space-time becomes a nearly-trivial, flat,
compactified Milne geometry, which is locally Minkowski.\cite{crunch_bang} 
After the bounce, in
the Einstein-frame description we use here, the universe is in an
expanding phase where modes no longer exit the horizon. Super-horizon
modes are frozen, and sub-horizon modes are well described by the
adiabatic (WKB) approximation so that new gravitational waves are not
generated.

\section{The present-day spectrum, $\Delta h(k,\tau_{0})$}

To convert from the primordial spectrum to the present-day spectrum
$\Delta h(k,\tau_{0}) \equiv T_{h}(k)\Delta h(k,\tau_{r})$ we need to
know the transfer function, $T_{h}(k)$.  To approximate $T_{h}(k)$,
note that $\Delta h(k,\tau)$ is roughly time-independent outside the
horizon, and decays as $a^{-1}$ once a mode re-enters the horizon.
Therefore, the transfer function is $\sim 1/(1+z_{r})$ for modes
already inside the horizon at the onset of radiation domination
$\tau_{r}$, and $\sim 1/(1+z_{k})$ for modes that entered at red shift
$z_k$ between $\tau_{r}$ and $\tau_{0}$.  Using the fact that
$H\propto a^{-2}$ during radiation domination and $H\propto a^{-3/2}$
during matter domination (neglecting the change in $g_{\ast}$) we find
\begin{equation}
\label{transfer_approx}
T(k)\approx\left(\frac{k_{0}}{k}\right)^{2}\left[1+
\frac{k}{k_{eq}}+\frac{k^{2}}{k_{eq}k_{r}}\right]
\end{equation}
where $k_{0}\equiv a_{0}H_{0}$, $k_{eq}\equiv a_{eq}H_{eq}$,
$k_{r}\equiv a_{r}H_{r}$, and $k_{end}\equiv a_{end}|H_{end}|$ denote
the modes on the horizon today ($\tau_{0}$), at matter-radiation
equality ($\tau_{eq}$), at the start of radiation domination
($\tau_{r}$), and at the end of the ekpyrotic phase ($\tau_{end}$),
respectively.

\begin{figure}
\begin{center}
\includegraphics[width=3.0in]{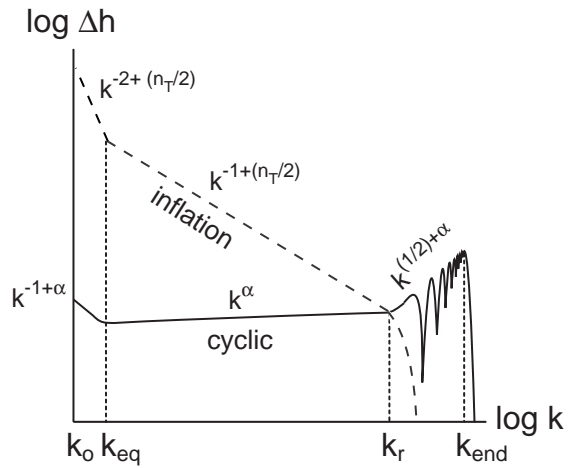}
\end{center}
\caption{A schematic comparison of the dimensionless strain observed
today $\Delta h(k,\tau_{0})$, as predicted by inflation and the cyclic
model. Here $n_{T}$ is the inflationary tensor spectral index (a small
negative number), and $\alpha\ll 1$ in the cyclic model is a small
positive number.  $k_{r}$ denotes the mode on the horizon at the start
of radiation domination.}
\label{schematic}
\end{figure}

The gravitational wave spectrum can be divided into three regimes.
There is a low frequency (LF) regime corresponding to long wavelength
modes that re-enter after matter-radiation equality ($k< k_{eq}$), and
a medium frequency (MF) regime consisting of modes which re-enter
between equality and the onset of radiation domination ($k_{eq}< k <
k_r$).  (We ignore the recent dark energy dominated phase, which has
negligible effect.)  The spectrum for these two regimes is:
\begin{equation}
\label{Gamma<1_spec}
\Delta h\approx\frac{\Gamma^{\frac{1}{2}}k_{0}^{2}}
 {\pi M_{pl}H_{r}^{\alpha}}\left\{\begin{array}{l}
 k^{-1+\alpha} \qquad (LF) \\
 k^{\alpha}/k_{eq} \qquad (MF)
\end{array}\right.
\end{equation}
Finally, modes which exit the horizon during the ekpyrotic phase
(before $\tau_{end}$), and re-enter during the expanding kinetic phase
(after the bound but before $\tau_r$) result in a high frequency (HF) band ($k_r< k<
k_{end}$):
\begin{equation}
\label{Gamma>1_spec}
\!\!\!\!\!\!
\Delta h\!\!\approx\!
\left(\!\frac{\!\sqrt{2}}{\pi}\!\right)^{\!\!\frac{3}{2}}
\!\!\frac{(\Gamma H_{r})^{\!\frac{1}{2}-\alpha}k_{0}^{2}}{M_{pl}k_{eq}k_{r}}
 \!\left|{\rm cos}\!\left(k\tau_{r}\!\!-\!\frac{\pi}{4}\right)\!\right|
 \! k^{\frac{1}{2}+\alpha}\quad\!\! (HF)
\end{equation}
The HF band runs over a range $k_{end}/k_r=\Gamma$, and this
quantity is strongly constrained by the requirement that the scalar
field cross the negative region of the potential before radiation
domination begins, which requires that\cite{cyclic_constraints}
\begin{equation}
H_r \lesssim  {V_{end}^{1\over 2}\over M_{Pl}}
\left({V_0\over V_{end}}\right)^{\sqrt{3\over2}/c},
\end{equation}
where $V_0$ is today's value of the dark energy density.  This
equation, combined with (\ref{def_Gamma}), gives a lower bound on
$\Gamma$, $\Gamma \gtrsim (V_{end}/V_0)^{\sqrt{2/3c^2}}$.  
For example,
for $V_{end}$ around the GUT scale and $c =10$, we find $\Gamma \ge
10^8$.  Fig.~\ref{schematic} schematically depicts $\Delta
h(k,\tau_{0})$ in the cyclic scenario and compares it to the
inflationary spectrum. \cite{Turner}

Another useful quantity is $\Omega_{gw}(k,\tau_{0})$, the
gravitational wave energy per unit logarithmic wavenumber, in units of
the critical density \cite{Thorne,Turner}:
\begin{equation}
\label{energy_strain_relation}
\Omega_{gw}(k,\tau_{0})\equiv\frac{k}{\rho_{cr}}\frac{d\rho_{gw}}{dk}
=\frac{1}{6}\left(\frac{k}{k_{0}}\right)^{2}\Delta h(k,\tau_{0})^{2}\;.
\end{equation}
In the cyclic model, $\Omega_{gw}(k,\tau_{0})$ is very blue, with
nearly all the gravitational wave energy concentrated at the
high-frequency end of the distribution.

\section{Observational constraints and Detectability}

The strongest observational constraint on the gravitational spectrum
in the cyclic model comes from the requirement that the successful
predictions of big bang nucleosynthesis (BBN) not be affected, which
requires
\begin{equation}
\label{BBN_constraint}
\int_{k_{BBN}}^{k_{end}}\Omega_{gw}(k,\tau_{0})\frac{dk}{k}
 \lesssim \frac{0.1}{1+z_{eq}}\;\;.
\end{equation}
From the above equations, (\ref{Gamma>1_spec}) and
(\ref{energy_strain_relation}), and using $1+z_{eq} \approx
k_{eq}^2/k_0^2$, and $T_r \sim H_r^{1\over 2}M_{Pl}^{1\over 2}$ for
the temperature at radiation domination, we obtain a total $\Omega$ in
gravitational waves of $\sim(2\alpha V_{end}/T_r M_{Pl}^3)^{4\over 3}
[36\pi^{3}(1+z_{eq})]^{-1}$, which from (\ref{BBN_constraint}) implies
\begin{equation}
\label{density_pert_constraint}
T_r \gtrsim {\alpha\over 20} V_{end} M_{Pl}^{-3},
\end{equation}
where, for simplicity, we have ignored the factor which
depends on the number of thermal degrees of freedom,
 which further weaken this bound.

The other observational constraints are much
weaker. \cite{Allen_Thorne} From the CMB anisotropy, one infers
$\Delta h(f\sim 10^{-18}\textrm{Hz})\lesssim 10^{-5}$; from precision
pulsar timing, $\Delta h(f\sim 10^{-8}\textrm{Hz})\lesssim 10^{-14}$.
Optimistic goals for LISA and advanced LIGO are strain sensitivities
of $\Delta h(f\sim 10^{-4}\textrm{Hz})\sim 10^{-20.5}$ and $\Delta
h(f\sim 10^{2} \textrm{Hz})\sim 10^{-24}$, respectively.  
 Fig.~3 shows results
for  values of $T_{r}$
and $V_{end}$ consistent with  all constraints on
the cyclic model  \cite{cyclic_constraints}.

Even if the  parameters are chosen to 
 saturate the BBN constraint, the spectrum
is still orders of magnitude
below the  sensitivity of anticipated instruments.
  In particular, it is hopeless to search
for an imprint in the polarization of the CMB from cyclic model
gravity waves because  the predicted amplitude on large scales is so small.
Hence, the detection of
a stochastic gravitational wave imprint in the CMB polarization would
be consistent with inflation and definitively rule out the cyclic
model.

\begin{figure}                                           
\begin{center}                                                                 
\includegraphics[width=3.0in]{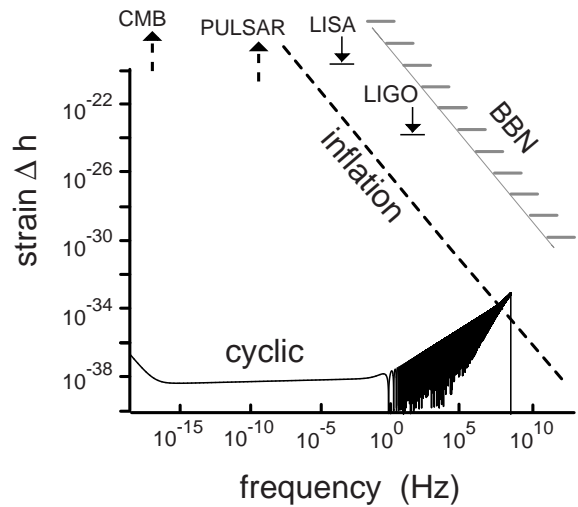}
\end{center}                                                                   
\caption{The present-day dimensionless strain, $\Delta h(k,\tau_{0})$,         
predicted by the cyclic model with $T_{r}= 10^{7}$~GeV and                     
$V_{end}^{1/4} = 10^{14}$~GeV. These parameters yield a gravity wave           
density four orders of magnitude below the BBN bound.  Some                    
observational bounds and (optimistic) future strain sensitivities are          
indicated. The dashed arrows mean the empirical bounds lie well above          
range of $\Delta h$ displayed here.}                                           
\label{real}                                                                   
\end{figure}          

We thank A. Tolley and J. Khoury for helpful conversations.  LB is
supported by an NSF Graduate Fellowship.  This work was also supported
in part by US Department of Energy grant DE-FG02-91ER40671 (PJS) and
by PPARC-UK (NT).

\end{document}